# THE STATISTICS OF NEARLY ON-AXIS GRAVITATIONAL LENSING EVENTS*

YUN WANG
*NASA/Fermilab Astrophysics Center*
*Fermi National Accelerator Laboratory, Batavia, IL 60510-0500*

**Abstract.** A small volume of space, nearly on-axis behind a gravitational lens with respect to a given source, will receive a greatly increased radiation flux. In the idealized case of a point mass lens acting on a point source in complete isolation, the volume will approach zero only as the flux tends to infinity; in fact, the volume weighted rms flux is divergent. In realistic cases, finite source size and the effects of other gravitational deflections (i.e., non-zero shear) limit the maximum flux and considerably complicate the physics, but very large fluxes are still produced in small volumes. We consider the physics and statistics of these Extreme Gravitational Lensing Events (EGLE) and present an initial examination of their possible astrophysical effects for various known and putative populations of lensing objects and sources, with particular attention to the case in which finite source size is important but shear is not.

## 1. Introduction

Nearly on-axis gravitational lensing events are extreme in magnification, and rare in occurrence for a given observer. However, every astrophysical object is a gravitational lens, as well as a receiver/observer of the light from sources lensed by other objects in its neighborhood. Statistically, extreme gravitational lensing events (EGLE) can have an effect on certain fragile objects in the sky, such as interstellar medium, molecular clouds, atoms, dust grains, etc.

An EGLE occurs when a moving observer crosses the line connecting a source and a lens. The maximum magnification of the source seen by the observer can be very large, limited only by the source size and the shear on the lens. As the observer moves away from the line connecting the source and the lens, the magnification of the source decreases. The duration of an EGLE depends on the velocity of the observer. A slowly moving observer in the neighborhood of a pair of small-size source and slightly-sheared lens can experience a strong burst of radiation due to the lensing of the source, which can be sufficient to affect the observer's properties.

If the observer moves a distance $d$ away from the line connecting the source and the lens, it is equivalent to the source moving an angular distance of $y$ (in units of the angular Einstein radius $\theta_{\rm E}$) from the optical axis (the line connecting the lens and the observer). For sufficiently distant lens or source, we have

$$y \simeq \left(\frac{D_{\rm ds}}{D_{\rm d}}\right) \frac{d}{D_{\rm s}\theta_{\rm E}}, \tag{1}$$

where $D_{\rm ds}$, $D_{\rm s}$, and $D_{\rm d}$ are angular diameter distances between the lens and source, observer and source, observer and lens respectively. The angular Einstein radius

$$\theta_{\rm E} = \sqrt{\frac{2R_{\rm S}D_{\rm ds}}{D_{\rm d}D_{\rm s}}} = 10^{-6} \times \sqrt{\left(\frac{M_{\rm L}}{5\times 10^6 M_\odot}\right)\left(\frac{1\,{\rm Mpc}}{D_{\rm d}}\right)\frac{D_{\rm ds}}{D_{\rm s}}}, \tag{2}$$

where $M_{\rm L}$ is the mass of the lens. The dimensionless size of a source with physical radius $\rho$ is defined as

$$R \equiv \frac{\rho}{D_{\rm s}\theta_{\rm E}} = \left(\frac{\rho}{3.09\times 10^{18}\,{\rm cm}}\right)\left(\frac{10^{-6}}{\theta_{\rm E}}\right)\left(\frac{1\,{\rm Mpc}}{D_{\rm s}}\right). \tag{3}$$

For a given pair of lens and source, the shear on the lens $\gamma \sim \tau$, (Wang and Turner, 1995) where $\tau$ is the optical depth for microlensing, the probability that the source is lensed. The presence of shear considerably complicates the physics. Since $\tau = \Omega_{\rm L} z_{\rm Q}^2/4$, (Turner, 1980; Turner *et al.*, 1984) where $\Omega_{\rm L}$ is the density fraction in lenses and $z_{\rm Q}$ is the redshift of small sources, the shear $\gamma$ is probably not important in the low redshift Universe. Here we briefly discuss the case in which the finite size of the source is important but shear on the lens is not, i.e., $\gamma < R$; a detailed discussion is presented in Wang and Turner (1995). For simplification, we only consider high magnification events (i.e., $R \leq 0.05$, $y \ll 1$).

## 2. Point source

Let us first consider a point source S with luminosity $L_{\rm S}$, being lensed by a lens L with Schwartzschild radius $R_{\rm S}$ (mass $M_{\rm L}$) at a distance $D_{\rm ds}$. In a narrow tube-shaped volume $V_{\rm SL}(f)$ behind the lens, which extends from the lens and tapers off to infinity, the flux from the source exceeds $f$. The cross section of the tube is $\sigma(f)=\pi d^2$. Hence

$$V_{\rm SL}(f)=\int_0^{D_{\rm d}(q)} {\rm d}D_{\rm d}\,\sigma(f). \qquad (4)$$

Summing over S gives the total volume $V_{\rm L}$ in which the flux exceeds $f$ for a given lens L; further summing over L gives the total volume $V_{\rm tot}(>f)$. We use $D_{\rm s}=D_{\rm d}+D_{\rm ds}$ for simplicity in our calculations.

In the absence of magnification, the flux from the source is $f_0=L_{\rm S}/(4\pi D_{\rm s}^2)$. The magnified flux $f=\mu f_0$. We find

$$V_{\rm SL}(f)=\frac{\pi R_{\rm S}}{D_{\rm ds}^2}\left(\frac{L_{\rm S}}{4\pi f}\right)^2, \qquad V_{\rm L}(f)=4\pi^2 n_{\rm S} R_{\rm S} D_{\rm c}\left(\frac{L_{\rm S}}{4\pi f}\right)^2, \qquad (5)$$

where $n_{\rm S}$ is the number density of sources, and $D_{\rm c}$ is the maximum separation between a lens and a source.

Let $\mathcal{F}_{\rm L}(f)$ be the volume fraction of space in which the flux from the source exceeds $f$ due to gravitational lensing, and $\mathcal{F}_{\rm S}(f)$ the volume fraction of space in which the flux from the source exceeds $f$ due to being close to the source. We have (Wang and Turner, 1995)

$$\mathcal{F}_{\rm L}(f)=n_{\rm L}V_{\rm L}(f)=\frac{3}{2}\tau N_{\rm S}\left(\frac{f}{f_{\rm min}}\right)^{-2}, \qquad \mathcal{F}_{\rm S}(f)=N_{\rm S}\left(\frac{f}{f_{\rm min}}\right)^{-3/2},$$
$$\frac{\mathcal{F}_{\rm L}(f)}{\mathcal{F}_{\rm S}(f)}=\frac{3}{2}\tau\left(\frac{f}{f_{\rm min}}\right)^{-1/2}, \qquad (6)$$

where $\tau$ is the optical depth, $N_{\rm S}$ is the total number of sources, and $f_{\rm min}=L_{\rm S}/(4\pi D_{\rm c}^2)$. Note that the volume weighted rms flux due to lensing diverges logarithmically.

## 3. Finite source

Let us now consider a source S with physical radius $\rho$ and luminosity $L_{\rm S}$, being lensed by a lens L with Schwartzschild radius $R_{\rm S}$ (mass $M_{\rm L}$) at a distance $D_{\rm ds}$. The tube-shaped volume $V_{\rm SL}(f,\rho)$ behind the lens in which the flux from the source exceeds $f$ has finite length $D_{\rm d}^{\rm m}(f)$, because of the finite size of the source.

For a finite source with dimensionless size $R$, $\mu_{\rm max}=2/R$. Therefore $f=\mu f_0\le\mu_{\rm max}f_0$. Let $f=\mu_{\rm max}f_0$ at $D_{\rm d}=D_{\rm d}^{\rm m}(f)$, i.e., the flux is equal to $f$ on the line connecting the source and the lens. For given $D_{\rm d}$, the flux decreases away from the line connecting the source and the lens, hence the volume in

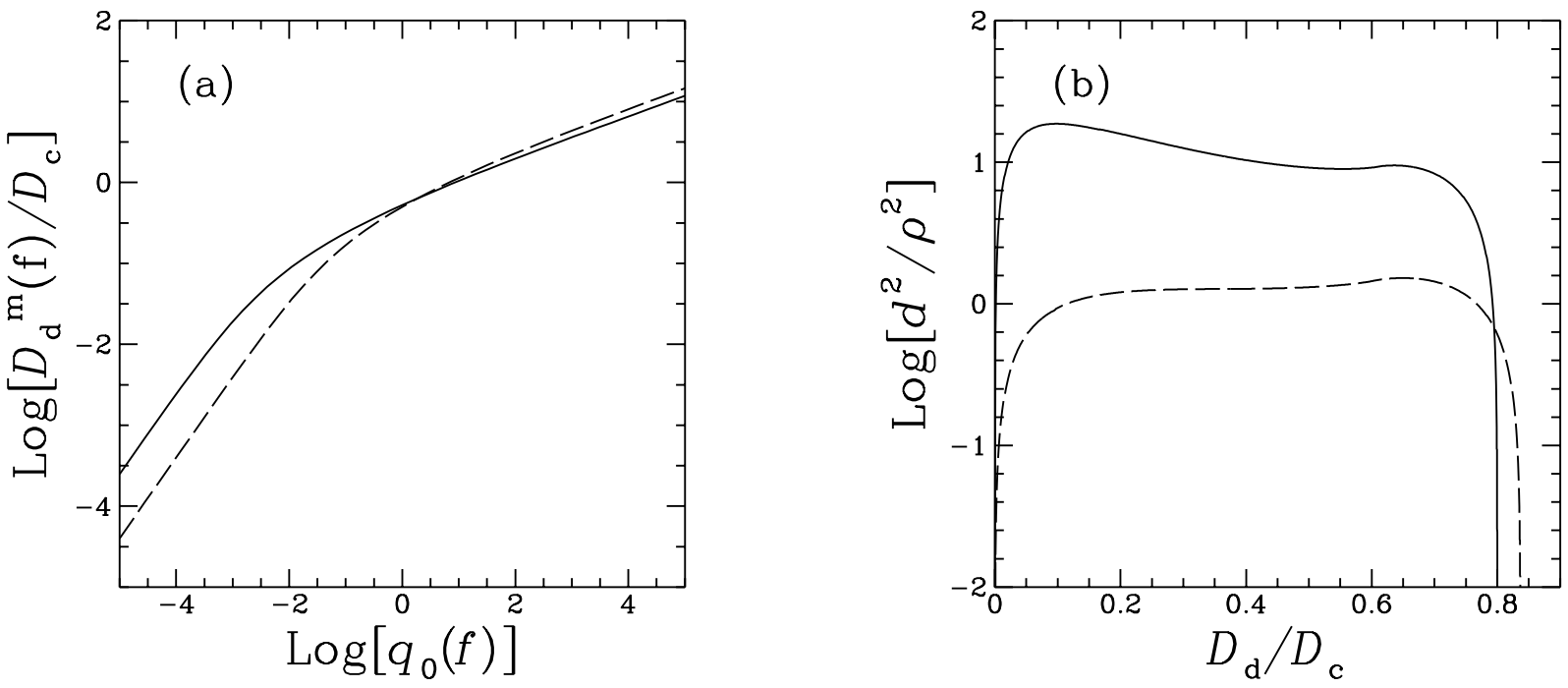

*Figure 1.* (a) Length of the tube volume $V_{\rm SL}(f,\rho)$, $D_{\rm d}^{\rm m}(f)/D_{\rm c}$; (b) Cross section of the tube volume $V_{\rm SL}(f,\rho)$.

which the flux exceeds $f$ converges to a point at $D_{\rm d}=D_{\rm d}^{\rm m}(f)$, i.e., $D_{\rm d}^{\rm m}(f)$ gives the length of the tube volume in which the flux exceeds $f$. $D_{\rm d}^{\rm m}(f)$ can be found analytically for arbitrary source size. (Wang and Turner, 1995)

Let us define a parameter $q_0(f)$ which measures the maximum magnification of the source relative to the minimum flux $f$,

$$q_0 \equiv \frac{8R_{\rm S}D_{\rm c}}{\rho^2}\left(\frac{L_{\rm S}}{4\pi D_{\rm c}^2 f}\right)^2. \tag{7}$$

Figure 1(a) shows the length of the tube volume behind the lens in which the flux exceeds $f$, $D_{\rm d}^{\rm m}(f)/D_{\rm c}$, as function of $q_0(f)$, for $D_{\rm ds}=0.2\,D_{\rm c}$ (solid line), $0.5\,D_{\rm c}$ (dashed line). The tube length is of order $D_{\rm c}$ for $q_0$ of order 1. Defining $x\equiv D_{\rm ds}/D_{\rm c}$, we have

$$\frac{D_{\rm d}^{\rm m}(f)}{D_{\rm c}} = \begin{cases} q_0(f)/x^2 & \text{for } q_0^{1/3} \ll x \\ [q_0(f)\,x]^{1/4} & \text{for } q_0^{1/3} \gg x \end{cases} \tag{8}$$

The tube volume $V_{\rm L}(f,\rho)$ has the cross-section $\sigma(f,\rho,D_{\rm d})$ which vanishes at $D_{\rm d}=0$, $D_{\rm d}^{\rm m}$. We can write

$$\sigma(f,\rho,D_{\rm d}) = \pi\rho^2 \times d^2/\rho^2 = \pi\rho^2\,\overline{\sigma}(q_0,D_{\rm d}), \tag{9}$$

for given $D_{\rm ds}$. Figure 1(b) shows the cross-section $\overline{\sigma}(q_0,D_{\rm d})=d^2/\rho^2$ with $q_0(f)=4$, for $D_{\rm ds}=0.2\,D_{\rm c}$ (solid line), $0.5\,D_{\rm c}$ (dashed line). The lens which is closest to the source has the thickest tube of high flux behind it.

The volume fraction in which the flux from the source exceeds $f$ is

$$\mathcal{F}_{\rm L}(f,\rho) = 4\pi^2 n_{\rm L} n_{\rm S} R_{\rm S} D_{\rm c}\left(\frac{L_{\rm S}}{4\pi f}\right)^2 I(q_0), \tag{10}$$

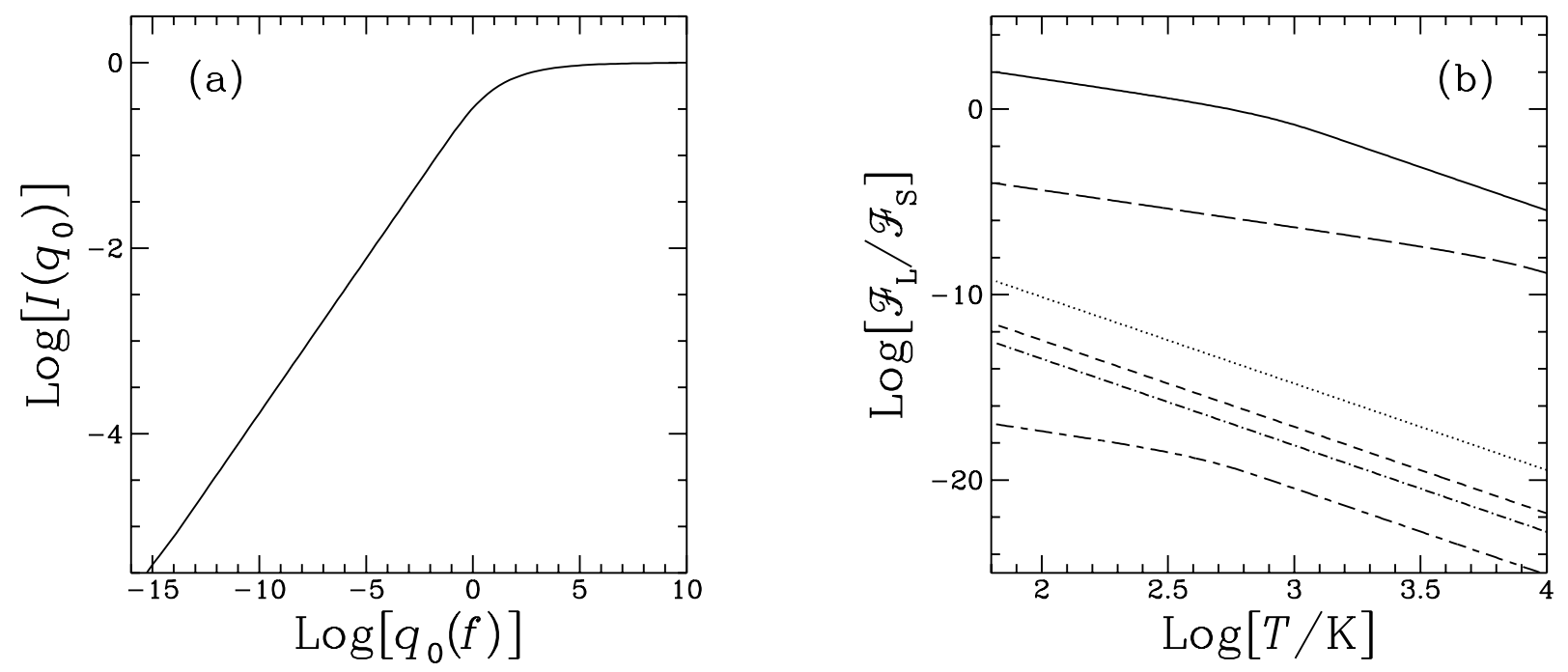

*Figure 2.* (a) Ratio of volume fractions, $I(q_0) = \mathcal{F}_L(f,\rho)/\mathcal{F}_L(f,\rho=0)$; (b) Volume fractions for different sets of sources and lenses.

where $I(q_0) = \mathcal{F}_L(f,\rho)/\mathcal{F}_L(f,\rho=0)$. To estimate $I(q_0)$, we can approximate the magnification of a finite source with the point source magnification cut off at $\mu_{\rm max} = 2/R$; this leads to a slight underestimate of $I(q_0)$, but it is sufficiently accurate for all practical purposes. $I(q_0)$ is plotted in Figure 2(a). For $q_0(f) \ll 1$, $I(q_0) = 0.3583\, q_0^{1/3}$; for $q_0 \gg 1$, $I(q_0) \simeq 1$. (Wang and Turner, 1995)

## 4. Possible astrophysical effects

In the context of EGLE, the relevant dimensional physical quantities are: lens mass $M_{\rm L}$, size of the lens-source distribution $D_{\rm c}$, source size $\rho$, source luminosity $L_{\rm S}$, minimum lensed flux from the source $f$. All these collapse into a single dimensionless parameter $q_0(f)$ (see Eq.(7)), which can be written as

$$\log(q_0) = -5.62 + 0.8\, m_{\rm bol} + 2\,\log\left(\frac{L_{\rm S}}{L_\odot}\right) + \log\left(\frac{M_{\rm L}}{M_\odot}\right) - 2\,\log\left(\frac{\rho}{R_\odot}\right) - 3\,\log\left(\frac{D_{\rm c}}{1\,{\rm kpc}}\right) \quad (11)$$

where the minimum flux $f$ is measured by $m_{\rm bol}$. For gamma-ray bursts, $\log(q_0) = 26 + 0.8\, m_{\rm bol}$ (lensed by stars), and $\log(q_0) = 34 + 0.8\, m_{\rm bol}$ (lensed by giant black holes).

Let us consider a population of sources (with number density $n_{\rm S}$) lensed by a population of lenses (with number density $n_{\rm L}$). For a given source, it induces one high flux tube behind each lens; for a given lens, it has one high flux tube coming out of it because of each source.

Let $\mathcal{F}_L(f)$ and $\mathcal{F}_S(f)$ denote the total volume fractions of space in which the flux from the source exceeds $f$ due to gravitational lensing and due to being close to the source respectively. We have (Wang and Turner, 1995)

$$
\begin{aligned}
\log \mathcal{F}_{\rm L} &= -8 + \log I(q_0) + 0.8 m_{\rm bol} + 2\log\left(\frac{L_{\rm S}}{L_\odot}\right) + \log\left(\frac{M_{\rm L}}{M_\odot}\right) \\
&\quad + \log\left(\frac{D_{\rm c}}{1{\rm kpc}}\right) + \log\left(\frac{n_{\rm S}}{1{\rm pc}^{-3}}\right) + \log\left(\frac{n_{\rm L}}{1{\rm pc}^{-3}}\right) \\
\log\left(\frac{\mathcal{F}_{\rm L}}{\mathcal{F}_{\rm S}}\right) &= -8.86 + \log I(q_0) + 0.2 m_{\rm bol} + \frac{1}{2}\log\left(\frac{L_{\rm S}}{L_\odot}\right) + \log\left(\frac{M_{\rm L}}{M_\odot}\right) \\
&\quad + \log\left(\frac{D_{\rm c}}{1{\rm kpc}}\right) + \log\left(\frac{n_{\rm L}}{1{\rm pc}^{-3}}\right) \qquad (12)
\end{aligned}
$$

where the minimum flux $f$ is measured by $m_{\rm bol} = -10\,\log(T/0.56{\rm K})$; $T$ is the minimum blackbody temperature of dust grains in the tube volume $V_{\rm SL}(>f)$. In Figure 2(b), we show $\mathcal{F}_{\rm L}/\mathcal{F}_{\rm S}$ versus $T$ for gamma-ray bursts lensed by stars (solid line) and by giant black holes (long dashed line), QSO's (X-ray) lensed by stars (dotted line) and by giant black holes (dot-short dashed line), neutron stars lensed by stars (short dashed line) and by giant black holes (short-long dashed line). Only orders of magnitude are used for the sources and lenses in Figure 2(b). The largest effect comes from gamma-ray bursts lensed by stars in our simple model (zero shear).

Finally, we note that although the volume fractions of high flux due to lensing are small, the corresponding absolute volumes can be large. Further, since materials move across the high flux tubes constantly, the fraction of material affected by EGLE is much higher than the static volume fractions. We are still investigating various possible astrophysical applications.

## 5. Acknowledgements

This work is done in collaboration with Ed Turner. The author is supported by the DOE and NASA under Grant NAG5-2788.